# Electromagnetic way of accelerating the magnetic dipoles

*S. N. Dolya*

*Joint Institute for Nuclear Research, Joliot - Curie str. 6, Dubna, Russia, 141980*

**Abstract**

The article considers an opportunity of electrodynamics accelerating the magnetic dipoles at initial velocity $V_{in}$ = 0.6 km / s, which is the magnetic dipole gain after pre-gas-dynamic acceleration to finite velocity $V_{fin}$ = 8.5 km / s. The acceleration length $L_{acc}$ = 2.27 km. The dipoles are accelerated at the forefront of the current pulse running through the spiral up. The accelerated magnetic dipoles have mass m = 1 kg, diameter $d_{sh}$ = 20 mm, overall length $l_{tot}$ = 65 cm, the length of the conical part of the $l_{cone}$ = 20 cm. When selecting the drag coefficient $C_x$ and the lift coefficient $C_y$ equal to ~ $10^{-2}$, the dipoles rise to height H = 10 km during a period of time τ = 14 s, thus reaching the vertical velocity $V_r$ = 1 km / s and reducing the forward velocity till $V_\varphi$ = 7.5 km / s. The magnetic dipoles reach flight range $S_{max}$ = 12300 km.

**Introduction**

There is a known [1] method of accelerating the magnetic dipole - the Gauss gun - a version of the electromagnetic mass accelerator, named after the German scientist Carl Gauss, who laid the foundations of the mathematical theory of electromagnetism. The Gauss gun consists of a solenoid with a trunk inside (usually dielectric). One end of the stem is inserted into the shell (made of ferromagnetic material). When the electric current flows in the solenoid the magnetic field which appears there, accelerates the projectile, "pulling" it into the solenoid. To keep the solenoid shell against being involved into the opposite direction, i.e., not to slow down, the solenoid must be turned off at this moment. That is why for the greatest effect the current pulse in the solenoid must be short-lived and powerful. As a rule, electrical capacitors with high-voltage are used to obtain this pulse.

However, despite the apparent simplicity of the Gauss gun and obvious advantages, its practical use is accompanied with serious complications.

Among ferromagnetic materials being used in the magnetic dipoles, iron is the most suitable one because of its high specific magnetic moment and high Curie temperature. The specific magnetic moment is a property of the useable substances and can not be increased. Moreover, due to the fact that the magnetic dipole must also include the jet engine with the fuel and navigation devices, the specific magnetic dipole moment will be even less than that of the pure iron. It does not allow the magnetic dipole in the Gauss gun to reach the high finite velocity.

To increase the finite velocity of the magnetic dipole, it is possible by two ways: to increase the amplitude of the accelerating current pulse or – the specific magnetic moment. A significant increase of the amplitude of the pulse causes a strong need to increase the dielectric durability of the insulation. The superconducting coil located inside the magnetic dipole increases the specific magnetic moment by several times in comparison with a pure iron dipole. This can significantly increase the acceleration rate and finite velocity of the magnetic dipoles.

To achieve the speed of 8.5 km /s, it is necessary to increase the length of the acceleration by a few kilometers, so that the accelerator will have to be placed horizontally. The corresponding angle between the direction of the velocity and the horizon, which is needed to go up through the atmosphere, must be installed by the lift force acting on the magnetic dipole. This means that the head part of the magnetic dipole must have the corresponding asymmetry, and the magnetic dipole should have the parts, stabilizing its orientation in space.

**Accelerator**

*1. Pre-acceleration of the magnetic dipoles with the gas-dynamic method*

To speed up the magnetic dipoles by the field of the running wave, the wave must be very slow. It must be mentioned that the relative velocity $\beta = 10^{-6}$, where $\beta = V / c$, $c = 3 * 10^5$ km / s - the speed of light in vacuum, consistent with the normal speed equal to: $V = 0.3$ km / s, and it is less than the speed of the sound in the air. The gas-dynamic acceleration method does not allow one to reach the speed significantly higher than the speed of the sound in the air. For example, specifications of the gun AP 35/1000 produced by the German company "Rheinmetall" are as follows: the initial rate of shooting $V_{in} = 1.5$ km / s, the diameter of the projectile: $d_{sh} = 35$ mm. The company "Mauser" is developing an aircraft gun with a caliber (diameter of the projectile) $d_{sh} = 30 - 35$ mm and the initial projectile velocity $V_{in} = 1.8$ km / s.

To achieve a low drag coefficient of the magnetic dipole, it is required to have a form of the dipole to be a cylinder with a pointed cone at the head part. At the same time, due to a small diameter of the magnetic dipole and its great length, it will be difficult to achieve the initial velocity $V_{in}$ like in the aircraft guns, therefore, in this case, the initial velocity of the magnetic dipole should be smaller than $V_{in} = 1.5$ km / s.

Probably, the use of the sabot projectile would allow one to achieve a greater initial velocity than we suppose.

*2. Selection of the basic parameters*

Let us choose the parameters of the accelerated dipole: $d_{sh} = 20$ mm, diameter of the total length of the dipole $l_{tot} = 65$ cm, the length of the conical part $l_{cone} = 20$ cm. The magnetic dipole is made of iron, the initial velocity of the magnetic dipole $V_{in} = 0.6$ km / s, the finite velocity of the magnetic dipole $V_{fin} = 8.5$ km / s, the gradient of the magnetic field in the pulse accelerating the magnetic dipole is taken as $\partial H_{zw} / \partial z = G = 2$ kGs / cm.

The magnetic moment per atom in iron, [2], page 524, is of the value $m_{Fe} = 2.219$ Bohr magneton. The table value of the Bohr magneton is: [2], page 31, $m_b = 9.27 * 10^{-21}$ erg / Gs. Taking into account that the atomic weight of iron $A_{Fe}$ is $A_{Fe} = 56$, we find that the magnetic moment per nucleon of the iron is: $m_{Fe} = 2 * 10^{-10}$ eV / (Gs * nucleon). The specific magnetic dipole moment may be increased if to place inside the dipole a superconducting coil of $Nb_3Sn$ and let the ring current go via it.

*3. The opportunity of increasing the specific magnetic moment in a magnetic dipole*

We calculate how much the specific magnetic moment - the magnetic moment per unit of the magnetic dipole mass – will grow, if in its cylindrical part with a length equal to $l_{cyl} = 40$ cm, there is a superconducting layer of $Nb_3Sn$ with radius $r_{cyl} = 1$ cm and thickness $\delta_{cyl} = 0.2$ cm . We assume that the current density in the superconductor is equal to, [2], page 312, $j_{sc} = 3 \times 10^5$ A/cm$^2$. Then linear density $J_{sc}$ of current (A / cm) in the superconducting layer will be $J_{sc}$ (A / cm) = $j_{sc} * \delta_{cyl} = 6 * 10^4$ A / cm. This linear current density on the surface of the superconductor will create the magnetic field $H_{sc}$ (kGs) = $1.226 * J$ (A / cm) ≈ 70 kGs, which does not contradict the opportunity of achieving the current density $j_{sc} = 3 * 10^5$ A/cm$^2$, [2], page 312.

The total current flowing in the superconducting layer is equal to: $I_{sc} = J_{sc} * l_{cyl} = 2.4 * 10^6$ A, and will cause the magnetic moment equal to

$M_{sc} = I_{sc} * \pi r_{cyl}^2 = 7 * 10^6 A * cm^2$ or in the system CGS $M_{sc} = 7 * 10^5$ erg / Gs.

The total mass of the superconducting layer can be calculated from the fact that the density of $Nb_3Sn$ superconductor is $\rho_{Nb3Sn} = 8$ g/cm$^3$, atomic mass A = 400, and the total superconductor $V_{sc} = 50$ cm$^3$ contains $N_{Nb3Sn} = 2.4 * 10^{26}$ nucleons. The specific magnetic moment, the magnetic moment per mass unit (nucleon), will be equal to the following:
$m_{sc} = M_{sc} / N_{Nb3Sn} = 2 * 10^{-9}$ eV / (Gs * nucleon), which is about 10 times higher than in iron, [2], page 524.

Let the mass of iron in the magnetic dipole be $m_{Fe} = 0.4$ kg, the mass of the superconductor is also $m_{Nb3Sn} = 0.4$ kg, the mass of the jet engine, fuel, instruments, navigation and control devices is $m_{fuel} = 0.2$ kg.
Then the specific magnetic moment in the magnetic dipole will be as follows:
$m_{md} = 8.8 * 10^{-10}$ eV / (Gs * nucleon), which is approximately by 4.4 times greater than in iron.

Let us take the gradient of the magnetic field accelerating the magnetic dipoles to be equal to: $\partial H_{zw} / \partial z = 2$ kGs / cm. In this case, the rate of the energy gain by the magnetic dipole will be:
$\Delta W_{sh} = m * \partial H_{zw} / \partial z = 1.76 * 10^{-6}$ eV / (cm * nucleon). To achieve the energy gain from the initial to the finite energy $W_{fin}$, corresponding to the finite velocity, $V_{fin} = 8.5$ km / s, $W_{fin} = 0.4$ eV / nucleon, the acceleration length $L_{acc} = W_{fin} / \Delta W_{sh} = 2.27$ km is required.

*4. Ways to achieve the desired parameters of the accelerator*

Now we come to defining the spiral where the acceleration of magnetic dipoles is expected to take place with a specific magnetic moment:
$m = 8.8 * 10^{-10}$ eV / Gs * nucleon from the initial velocity: $\beta_{in} = 2 * 10^{-6}$ to the finite velocity: $\beta_{fin} = 2.83 * 10^{-5}$, $\beta = V / c$, $c = 3 \times 10^{10}$ cm / s - velocity of light in vacuum.

The radii of the spiral, [3], the initial and finite ones are taken as follows: $r_{0in} = 50$ cm, $r_{0fin} = 30$ cm. The accelerator, due to damping, will be divided into sections. Therefore, within one section it is possible to choose the initial and finite values of the parameter $x = 2\pi r_0 / \beta \lambda_0$, close to the optimal ones and equal to $x = 6.28 * 50 / (2 * 10^{-6} * 1.3 * 10^8) = 1.2$.

Selection of wavelength $\lambda_0 = 1.3 * 10^8$ means that we have chosen the pulse duration to be equal to the following:
$(f_0 = c/\lambda_0 = 230$ Hz$)$, $\tau_{pulse} = 1 / (2f_0) = 2.17$ ms.

The slowdown wavelength is equal to the following:
$\beta\lambda_0 = 2 * 10^{-6} * 1.3 * 10^8 = 260$ cm, and the magnetic field gradient
$\partial H_{zw} / \partial z = G = 2$ kGs / cm, and corresponds to the amplitude of the magnetic field pulse: $H_{zw} = 82.8$ kGs. This amplitude of the magnetic field can be found from the relation $\partial H_{zw} / \partial z = k_3 H_{zw} = 2\pi H_{zw}/\beta\lambda_0$. From this we get the value $H_{zw} = \beta\lambda_0 * G/2\pi = 82.8$ kGs.

To find the power flux needed to generate the magnetic field of this strength, we find a relation between the components of the electric field
$E_{zw} = E_0 I_0 (k_1 r)$ and of the magnetic field:
$H_{zw} = (k_1 / k) tg\Psi I_0 (k_1 r_0) E_0 I_0 (k_1 r) / I_1 (k_1 r_0)$, [3]. For the interior of the spiral, where $k_1$ is the transverse wave vector: $k = (\omega / c) * \varepsilon^{1/2}$ - the wave vector, $r_0$ - the radius of the spiral, the expression is as follows: $(k_1 / k) = 1/\beta_{ph}$, $tg\Psi \approx h/2\pi r_0$. So, $(k_1 / k) * tg\Psi = \varepsilon^{1/2}$ at the beginning of the spiral $k_1 r_0 = 1.2$ and the ratio of $I_0 (k_1 r_0) / I_1 (k_1 r_0) = 2$. Thus, the component of the magnetic field $H_{zw} = 82.8$ kGs on the spiral axis that corresponds to the electric field strength: $E_{zw} \approx 347.2$ kV / cm.

The value in the brackets, { }, (in formulae (13) in [3]), for the value of the argument $x = 1.2$,
$\{\} = \{(1 + I_0 K_1/I_1 K_0) (I_1^2 - I_0 I_2) + \varepsilon (I_0/K_0)^2 (1 + I_1 K_0 / I_0 K_1) (K_0 K_2 - K_1^2)\}$ is:
$\{\} = 3{,}77 * \varepsilon$ so that the power required to achieve the electric field strength $E_{zw} = 347.2$ kV / cm for the initial speed of the magnetic dipole $\beta_{in} = 2 * 10^{-6}$, may be found from formula [3]:

$$P = (c / 8) * E_{zw}^2 * r_0^2 * \beta * \{\}. \qquad (1)$$

The wave power, in Watts, is equal to [3]:

$$P (W) = 3 * 10^{10} * (3.47)^2 * 10^{10} * 2.5 * 10^3 * 2 * 10^{-6} * 1.28 * 10^3 * 3.77 /$$
$$(8 * 9 * 10^4 * 10^7) = 12.23 \text{ GW}. \qquad (2)$$

According to formula (2), to achieve the magnetic field gradient on the axis of the field $G = 2$ kGs / cm, it is required to have power $P = 12.23$ GW. This power can be achieved by using the pulse technology.

We introduce the notion of pulse amplitude $\tilde{U}_{acc}$ related with the field power on the axis of the spiral $E_{0pulse}$ by the following ratios [3]:

$$\tilde{U}_{acc} = E_{0pulse}\lambda_{slow}/2\pi, \quad \lambda_{slow} = \beta\lambda_0, \quad \lambda_0 = c/f_0. \qquad (3)$$

Thus, the amplitude of the power pulse propagating along the spiral must be equal: $\tilde{U}_{acc} = E_{0pulse} * \lambda_{slow}/2\pi = 14.37$ MV.

Table 1 summarizes the main parameters of the accelerator of magnetic dipoles.

Table1. Parameters of the accelerator.

| Parameter | Value |
| --- | --- |
| $m = 8.8*10^{-10}$, the dielectric outside spiral, $\partial H_z / \partial z = 2$ kGs / cm, wave power, P | P = 12.23 GW, $\mu = 1$, $\varepsilon = 1280$ |
| Velocity, initial –finite, $\beta_{ph}$ | $\beta_{ph} = 2*10^{-6} - 2.83*10^{-5}$ |
| The radius of the spiral, initial -finite, $r_0$ | $r_0 = 50 - 30$ cm |
| Frequency of the wave, $f_0$ | $f_0 = 230$ Hz |
| The electrical field strength, $E_{zw}$ | $E_{zw} = 347$ kV/cm |
| Accelerator length, $L_{acc}$ | $L_{acc} = 2.27$ km |
| Pulse duration, $\tau$ | $\tau = 2.17$ ms |
| The amplitude of the voltage $\tilde{U}_{acc}$ | $\tilde{U}_{acc} = 14.37$ MV |

*5. Preventing the turn of the dipole by $180^0$ in the magnetic field pulse by imposing the uniform magnetic field*

Everything would have been well if the dipole were a point. But the magnetic dipole is not a point and the action on it by the radial component of the magnetic field leads to "roll over", a reversal position of the dipole. Two interacting coils will seek to be placed in such a way that their planes should be parallel to each other, and the direction of the currents of the both would be the same. In contrast to the forces accelerating the dipole and leading to the radial displacement of the center of mass of the dipole, the action of the pair of the forces, leading to the "reversal" of the dipole, is summed up.

The simplest solution that prevents the turn of the dipole by $180^0$ in the accelerating field of the wave is the imposition of the uniform external magnetic field which won't influence on the acceleration of the dipole because of its homogeneity but will only hold the dipole against the reversal in the space.

Impose the external magnetic field $H_{out}$ on the spiral with the dipole to keep it against turning off. The magnetic field of the wave is of the order of $H_{zw} = 82.8$ kGs, respectively, to compensate small deviations $\sin\theta < 1$, the external magnetic field must be $H_{out} > 100$ kGs.

Thus, we have fixed the magnetic dipole orientation in the space. Now the reversal moment will act on the turns of the coils, which carry the impulse current, and therefore they must be reliably fixed. One can consider a mechanical model of such an accelerator, which should consist of two magnets directed to each other with the same sign poles. One of the magnets - is a magnetized iron magnetic dipole. The second magnet - is the current pulse running via the coils of the spiral. The one-sign poles of the magnets repel each other - the current pulse in the spiral is speeding up and accelerates (pushes) the magnetic dipole.

In this accelerator technology it is called the automatic phase stability principle [4]. Thus, the running current pulse can only push (not pull) the dipole being accelerated.

A dedicated narrow channel (trunk) holds the magnet being pushed against turning over. In our case the constant external magnetic field fulfills the role of this channel. The pushing magnet cannot "turn around" either; it is prevented from turning by mechanical forces.

*6. Radial focusing by the magnetic field having a variable component*

To eliminate the opportunity for the magnetic dipoles to "escape" along the radius, we have to introduce the radial focusing. The easiest way to do it is to impose an additional sinusoidal field over the external uniform field $H_{out}$, so that the total field will be a combination of the constant field $H_{out0}$ holding the magnetic dipole from the "turn around", and of the variable component of $H_{out1} * \sin k_z z$, which automatically appears when the solenoid coils producing the longitudinal magnetic field $H_{out}$, are positioned quite seldom. Thus, $k_z = 2\pi/a_z$, where $a_z$ is a spatial period of positioning the coils. This ripple, corrugated $H_{out} = H_{out0} + H_{out1} * \sin k_z z$ of the magnetic field leads to the appearance of sign-variable gradient of the external magnetic field and the appearance of sign-alternative forces in the equation of radial motion, assuming $k_\perp \approx k_z$,

$$r_1 \text{''} = \omega_r^2 r_1 + \tfrac{1}{2} W_{\lambda m} \sin k_z V_z t * [H_{out1} / (H_{out0} - H_{zw})] * k_z r_{b0} * (k_z c)^2 r_1, \quad (4)$$

where $k_z V_z$ - oscillation frequency in the external magnetic field.

We change variables $k_z V_z t = \omega_{out} t = 2\tau$, then $\partial / \partial t = \tfrac{1}{2} \omega_{out} \partial / \partial \tau$, and the equation (4) can be written as follows:

$$r_1 \text{''} = \{4 (\omega_r^2 / \omega_{out}^2) -$$

$$-2 W_{\lambda m} * ([H_{out1} / (H_{out0} - H_{zw})] * k_z r_{b0} * (k_z c)^2 / * \omega_{out}^2) * \sin 2\tau\} r_1, \quad (5)$$

and, thus, the equation is reduced to the Mathieu equation, [5], which describes the transverse motion:

$$r \text{''} + (a - 2q * \sin 2\tau) r = 0, \quad (6)$$

where

$$a = 4 (\omega_r / \omega_{out})^2, \quad (7)$$

$$q = W_{\lambda m} * [H_{out1} / (H_{out0} - H_{zw})] * k_z r_{b0} * (k_z c)^2 / \omega_{out}^2. \quad (8)$$

Depending on the values of parameters a and q, in the Mathieu equation there are areas of stable and unstable solutions. Choosing the alignment of the coils forming the magnetic field ($k_z$) and the current in them ($H_{out1}$) in such a way that the solution of (6) will be within the range of sustainability and it will be possible to obtain a stable acceleration of the magnetic dipoles. The stability condition is as follows [5]: $0.92 > q > a$. simplification by $W_{\lambda m}$, $k_z r_{b0}$, $\omega_{out}^2$ and $(H_{out0} - H_{zw})^{-1}$, we get

$$(k_z c)^2 H_{out1} > 8 (\pi f_0 / \beta_z)^2 * H_{zw}, \quad (9)$$

$$H_{out1} > 8 (k_3 / k_z) * (\pi f_0 / \beta_z k_z c)^2 H_{zw}, \quad (10)$$

$$H_{out1} > 4\pi (a_z / \lambda_{slow})^2 H_{zw}. \quad (11)$$

The amplitude of the alternating field must exceed the field of a wave multiplied by $4\pi (a_z / \lambda_{slow})^2$.

Perhaps, small focusing magnetic elements can be placed close to the axis of the spiral waveguide. Just as in the drift tube linear accelerator of Alvarez, the magnetic elements, maybe, won't greatly affect the distribution of the accelerating pulse propagating via the spiral waveguide. Then the spatial period of the magnetic field changes can be made small enough, $a_z \approx 10$ cm.

Substituting the previous expression with the values: $H_{zw} = 82.8$ kGs, $f_0 = 230$ Hz, $\beta_z = 2 * 10^{-6}$, $a_z \approx 10$ cm, $c = 3 \times 10^{10}$ cm / s - velocity of light in vacuum, we obtain:

$$H_{out1} \geq 1.85 * 10^{-2} * H_{zw} = 1.5 \text{ kGs}. \qquad (12)$$

This value of the variable component of the magnetic field will generate the sign-variable field on the basis of permanent magnets of NdFeB, similar to how it is done to focus the electron beams in traveling-wave tubes.

You can even eliminate the static magnetic field, but, nevertheless, hold the dipole against the reversal over the radius to carry out this only by the sign-variable magnetic field that has no DC component. This result is obtained if the amplitude of the external field $H_{out}$ is 2 times larger than the pulse field $H_{zw}$, multiplied by the ratio of half-cycle of the external field $a_z$ to the spatial length of the pulse $l_p$: $H_{out} > 2 (a_z / l_p) H_{zw}$.

### 7. The power damping of the pulse propagating in the spiral

Wave attenuation in a spiral waveguide will lead to the fact that the amplitude of the pulse traveling along the spiral will decrease while pulse moving from the beginning to the end of the spiral and this power reduction is related with the ohm currents for heating the spiral.

Current $I_\varphi$ flows through the windings of the spiral and, in fact, it is Ohm losses:

$$\Delta P \text{ (W / turn)} = \tfrac{1}{2} I_\varphi^2 * R, \qquad (13)$$

where $I_\varphi$ - the current going via the coil in amperes, R - loop resistance in ohms. Then $\Delta R$ / turn - is expressed in Watts.

We first find the resistance of the coil. The resistance is calculated by the usual formula: $R = \rho l / s$, where $\rho = 1.7 * 10^{-6}$ Ohm * cm - the resistivity of copper, the copper coil will be considered, $l = 2\pi r_0$ - coil length, $r_0$ - the radius of the spiral, s – the coil cross-section. Since the current flowing via the coil is of high-frequency (AC), a factor, ½ appears there and this current penetrates into the conductor for the depth of the skin – layer which has to be found.

The expression for the depth of the skin - the layer can be written as follows:

$$\delta = a / (\sqrt{2\pi\sigma\omega_0}), \qquad (14)$$

where $c = 3 \times 10^{10}$ cm / s - the speed of light in vacuum, $\sigma = 5.4 * 10^{17}$ 1 / s - the conductivity of copper, $\omega_0 = 2\pi f_0$ - circular frequency, $f_0 = 260$ Hz - frequency of the wave propagating in the spiral. Substituting numerical values in the formula (14) gives $\delta = 0.4$ cm.

The result is the depth of the skin - layer $\delta = 0.4$ cm, that is much larger than the distance between the turns of the spiral $h = 1 / n = 0.02$ cm, $n \approx 50$ - the number of turns of the spiral per 1 cm length of the spiral. This means that in order to reduce the resistance of one turn, and, accordingly, - reduce the attenuation, it is necessary to wind the coil as a rather wide tape with a width $H = 2\delta = 0.8$ cm. The tape should be placed on the wide side H over the radius. The distance h between the coils consists of h/2 turns and h/2 space between the coils, so that the winding pitch of the value h / 2 revolution, and the turn occupied the space h / 2, to be equal to the space between the coils.

Then the resistance of one coil $R = \rho l / s$ will be:

$$R = \rho * 2\pi r_0 / (2\delta * h / 2) = \rho * 2\pi r_0 / (\delta * h). \qquad (15)$$

Substituting numerical values for the start of spiral $r_0 = 50$ cm, we find

$$R = 1.7 * 10^{-6} * 6.28 * 50 / (2 * 0.4 * 10^{-2}) = 6.6 * 10^{-2} \text{ Ohm}. \qquad (16)$$

Now we have to find $I_\varphi$ - current in the coils. To do this, we use the formula

$$H_{zsurf} = (4\pi / c) nI_\varphi, \qquad (17)$$

where $H_{zsurf}$ - the magnetic field on the surface of the spiral, $H_{zsurf} \approx H_{zw}$. Now we can find current $nI_\varphi$ flowing via the coils of the spiral:
$nI_\varphi$ (A / cm) $= H_{zsurf} / (4\pi / c) = (1.226)^{-1} * H_{zsurf}$ (A / cm) $= H_{zsurf}$ (Gs).
Then – the current in one coil is:

$$I_\varphi \text{ (A)} = H_{zsurf} \text{ (Gs)} / n. \tag{18}$$

Substituting numerical values into the formula (18) we have found that the current in one coil is:
$I_\varphi$ (A) = [82.8 kA / cm] / (50 turns / cm) = 1.656 kA / revolution.

Ohm losses of the current in one coil are as follows:

$$\Delta P \text{ (W / turn)} = \tfrac{1}{2} I_\varphi^2 * R = 90 \text{ kW / coil}. \tag{19}$$

Since there are n turns per 1 cm, the energy losses per 1cm will be by n times more:

$$\Delta P \text{ (W / cm)} = \tfrac{1}{2} I_\varphi^2 * R * n = 4.5 * 10^6 \text{ W / cm}. \tag{20}$$

We introduce the ratio

$$\Delta P / P = -2\alpha, \tag{21}$$

hence,
$$1 / \alpha = L_{damping} = 2P/\Delta P = 2 * 12.23 * 10^9 / 4.5 * 10^6 =$$
$$= 5.43 * 10^3 \text{ cm} = 54 \text{ m}, \tag{22}$$

is the length at which the field power is reduced by a factor of e due to damping. It can be seen that the movement of the magnetic dipole while accelerating is necessary to calculate taking into account the damping of the power pulse propagating via the spiral waveguide.

*8. Magnetic dipole capture in the acceleration mode. Excesses*

We calculate the required accuracy of coincidence of the initial phase of the accelerating wave (pulse) with the synchronous phase. The theory of particle capture in a traveling wave gives, [6]: $\Delta\varphi = 3\varphi_s$, $(+ \varphi_s - 2\varphi_s)$. In practice, it means, for example, in our case, where $\lambda / 4$ corresponds to the duration of 1 ms or $90^0$, that one degree corresponds to the time interval of approximately 10 μs. In linear accelerators the buncher gives the phase width of the bunch $\pm 15^0$, and to avoid large phase fluctuations, we require that the synchronization accuracy of the magnetic dipole with the accelerating pulse was as follows:
$\Delta\tau = \pm 15 * 10$ μs $= \pm 150$ μs. Such synchronization accuracy seems to be quite reachable for the gunpowder start, i.e., - preliminary gas-dynamic acceleration of the magnetic dipoles.

Let us now calculate the excess for the accuracy of coincidences of the initial velocity of the magnetic dipole and the phase velocity of the pulse propagating along the spiral structure. We introduce value $g = (p-p_s) / p_s$ - the relative difference between the pulses [6]. In the non-relativistic case – it is just a relative velocity spread out of $g = (V-V_s) / V_s$. The vertical range of the separatrix is calculated with the following formula [6]:

$$g_{max} = \pm\, 2\, (\Omega_s/\omega_0)\, [1 - \varphi_s / ctg\varphi_s]^{½}, \qquad (23)$$

wherein: $\varphi_s = 45^0 = \pi / 4$, $ctg\varphi_s = 1$, $[1 - \varphi_s / ctg\varphi_s]^{1/2} = 0.46$, $2 * 0.46 = 0.9$
$\omega_0 = 2\pi f_0 = 1.44 * 10^3$, $\Omega_s/\omega_0 = [W_{\lambda m} ctg\varphi_s / 2\pi\beta_s]^{1/2}$.

Let us determine the value of $W_{\lambda m} = (m\, \partial\, Hz / \partial\, z)\, \lambda_0 sin\varphi_s/Mc^2$ - a set of the relative magnetic dipole energy at the wavelength $\lambda_0$ in vacuum. In our case $\lambda_0 = c/f_0 = 1.3 * 10^8$ cm, $sin\varphi_s = 0.7$, $Mc^2 = 1$ GeV, $W_{\lambda m} = 3.64 * 10^{-8}$, $ctg\varphi_s = 1$. Substituting numerical values we get $g = (V_{in}-V_s) / V_s = \Delta V / V_s$, and, finally, $\Delta V / V_s = \pm\, [3.64 * 10^{-8} / (6.28 * 10^{-6} * 2)]^{1/2} * 0.9 = \pm\, 0.05$.

Thus, the allowed discrepancy of the initial speed of the magnetic dipole with the pulse transfer is of the order of $\Delta V / V_s = \pm\, 5\%$.

### 9. *Magnetic dipoles from vacuum to atmosphere*

Magnetic dipoles should be accelerated in a rather high vacuum $P_1 \approx 10^{-6}$ mm. Hg. while their use is expected under normal atmospheric conditions: $P_2 \approx 10^3$ mm. Hg. So, the pressure difference is about nine orders of the magnitude. To reach such a pressure gradient, we can use several buffer cavities which are cylindrical chambers separated from each other by walls with the imbedded pulse iris diaphragms. Each cavity should be supplied with individual vacuum pumping.

We calculate the amount of the air particles which penetrate to the first cavity - the nearest one to the atmosphere. Let the radius of the iris be $r_{0d} = 10$ cm, and the time to open it $t_1 = 10^{-3}$ s. Then the linear velocity of the iris petals will be $V_{11} = r_{0d} / t_1 = 10^4$ cm / s, which should not cause a problem for the operation of the mechanism. The average velocity of the thermal motion of the molecules of air V is assumed roughly to be equal to the speed of sound in the air:

$V = 3 * 10^4$ cm / s. From the various spatial orientations of the velocity only 1/6 (1/6 - one facet of the cube) is directed towards the aperture. The number of molecules per cubic centimeter of air under normal conditions, the number of Loschmidt is $\rho_{0L} = 2.7 * 10^{19}$ molecule/cm$^3$. Then the number of molecules from the atmosphere which penetrated to the first buffer cavity when the diaphragm was open, will be as follows:

$$N_0 = (1/6) \rho_{0L} * \pi r_{0d}^2 * Vt_1. \qquad (24)$$

Substituting the numerical values into formula (24) we find that the total number of particles coming from the atmosphere to the first buffer cavity is: $N_0 = 4 * 10^{22}$ particles. Let the volume of the first cavity is the value equal to $V_{01} = 10^3 \, l = 10^6$ cm$^3$. Then the density of molecules in it after the diaphragm shutdown is equal to $n_0 = 4 * 10^{16}$ molecule/cm$^3$.

Particle density (and pressure: $p = nkT$) in the first buffer cavity is about 3 orders lower than the density of particles in the atmosphere under normal conditions. So, at least, three cavities of this volume will be required to reach the appropriate pressure gradient.

Now we consider the dynamics of the particle density in the cavity for the period of time between the shutdowns of the trigger device. Let the cavity be pumped out through the hole of the square of $S_1 = 10^4$ cm$^2$. We assume that all the molecules which got into this area are removed from the volume. It is supposed that the time between the cycles is $t_2 = 10^{-1}$ s, it means that the frequency of operation of the device is equal to $F = 10$ Hz. The equation describing the reduction of the density of the particles while pumping out can be written as follows:

$$dn = -n * (1/6) * S_1Vt/V_{01}. \qquad (25)$$

The solution of this equation can be written as

$$n = n_0 * \exp[- * (1/6) * S_1Vt/V_{01}]. \qquad (26)$$

For evacuation time $t = t_2$ the exponent is approximately equal to 5, thus, due to pumping out the density of molecules in the first buffer cavity is reduced by more than 100-times, $n = n_0 * 7 * 10^{-3}$. The density of particles in the first volume of the buffer before the next shutdown will be the value of $n_1 = 4 * 10^{16} * 7 * 3 * 10^{-3} = 10^{14}$ molecule/cm$^3$, that is by 5 orders of magnitude less than the number of Loschmidt: $\rho_{0L} = 2.7 * 10^{19}$ molecule/cm$^3$, corresponding to the number of molecules per cubic centimeter of the air under normal conditions.

It is evident that before the next shutdown the cavity can be considered to be empty.

Figure 1 shows a possible scheme of the device.

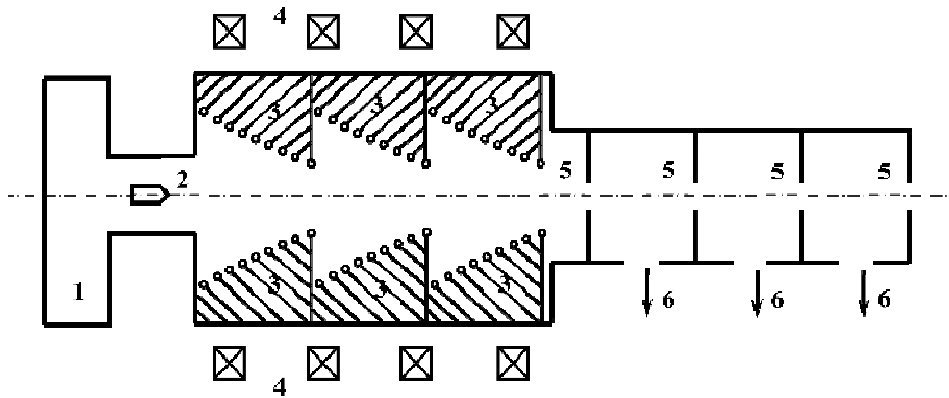

Fig.1. 1 – gun, 2 – dipoles, 3 - spiral section of the waveguide,
4 - current coils, 5 - pulse aperture, 6 - individual evacuation buffer cavities.

**Applications**

*1. Lifting power*

When the length of the accelerator $L_{acc} \approx 2.27$ km it can be positioned only horizontally. To transfer the magnetic dipole above the atmosphere, you can use a small asymmetry in the form of a magnetic dipole to create the lifting power $F_y$. The equation of the vertical motion in this case may be written as follows:

$$mdV_y / dt = C_y \rho_0 V_x^2 * S_{tr} / 2, \qquad (27)$$

where $C_y$ is an aerodynamic lift coefficient, $\rho_0 = 1.3 * 10^{-3}$ g/cm$^3$ - the air density on the surface of the Earth, $V_x = 8.5$ km / s - the horizontal speed of the magnetic dipole, $S_{tr}$ – the transversal cross-section of the magnetic dipole.

It is required that during $t_{fly} = 10$ s the magnetic dipole rises by a height of $H_{fly} = 10$ km, where the air resistance is negligible. From (27) we find the proper lifting coefficient $C_y$ for this case:

$$C_y = 4mH_{fly} / (\rho_0 V_x^2 * S_{tr} * t_{fly}^2). \qquad (28)$$

Substituting the numbers into the expression (28): $m = 10^3$ g, $S_{tr} = 3.14$ cm$^2$, we find that the lifting coefficient $C_y$ must be equal to $C_y \approx 10^{-2}$, that is, probably, not difficult to fulfill by a bent angle in the head part of the magnetic dipole.

*2. Ballistics. Air resistance*

It is necessary to calculate the motion of magnetic dipoles accelerated by using the electrodynamics method. The equation of motion of magnetic dipoles can be written as

$$mdV / dt = - \rho C_x S_{tr} V^2 / 2, \qquad (29)$$

where m is the mass of the magnetic dipole, V-velocity, $\rho = \rho_0 e^{-z/H0}$ - barometric formula of changing the atmospheric density with altitude, $\rho_0 = 1.3 * 10^{-3}$ g/cm$^3$ - the air density at the surface of the Earth, $H_0 = 7$ km - the altitude at which the density drops by factor e.

The aerodynamic coefficient or drag coefficient is called a dimensionless value that takes into account the "quality" of the form of the magnetic dipole:

$$C_x = F_x / (½) \rho_0 V_0^2 S_{tr}. \qquad (30)$$

The solution of equation (29) can be written as follows:

$$V(t) = V_0 / [1 + \rho C_x V_0 * S_{tr} * t/2m]. \qquad (31)$$

To calculate the change of the speed of magnetic dipoles, it is necessary to find the aerodynamic coefficient $C_x$.

*3. The calculation of the drag coefficient of magnetic dipoles for air*

It is assumed that the magnetic dipole has the form of a cylindrical rod with a conical head part. Then, at the hit of a nitrogen molecule on a sharp cone, the change of the longitudinal velocity of the molecules is equal to

$$\Delta V_x = V_x * \Theta_h^2 / 2, \tag{32}$$

where $\Theta_h$ is the angle at the vertex of the cone. The gas molecules transfer the momentum to the magnetic dipole:

$$p = mV = \rho V_x S_{tr} t * \Delta V_x. \tag{33}$$

The change in the momentum per unit of time is the power of the frontal inhibition-

$$F_{x1} = (½) \rho V_x S_{tr} * V_x * \Theta_h^2. \tag{34}$$

Dividing $F_{x1}$ by $(½) \rho V^2_x S_{tr}$, we get the drag coefficient for a sharp cone in the mirror reflection of the molecules from the cone (the Newton formula):

$$C_{x\ air} = \Theta_h^2. \tag{35}$$

Let the length of the cone part of the magnetic dipole is $l_{cone} = 20$ cm at diameter $d_{sh} = 20$ mm. This means that the angle at the vertex of the cone is $\Theta_t = 10^{-1}$ and $C_{x\ air} = 10^{-2}$.

In order to have a sharp cone in the head part of the magnetic dipole, it must be long enough. Limiting the length of the magnetic dipole means that for a good efficiency of its acceleration, the length of magnetic dipole $l_{tot}$ should be less than the slowdown wavelength quarter $\lambda_{slow} = \beta\lambda_0$, i.e.: $l_{tot} < \beta\lambda_0 / 4$. In our case, to start the acceleration, $\beta\lambda_0 / 4$ is $= 65$ cm.

*4. Passage of the magnetic dipoles through the atmosphere*

We set a table of the time dependence of the vertical velocity of the magnetic dipole, its lifting altitude and horizontal speed. The vertical velocity is calculated by the following formula:

$$\Delta V_y = C_y \rho V_x^2 * S_{tr} * \Delta t / 2m. \tag{36}$$

The altitude taking, respectively, is calculated as

$$H_{\text{fly } n+1} = H_{\text{fly } n} + \bar{V}_y * \Delta t + C_y \rho V_x^2 * S_{tr} * (\Delta t)^2 / 4m, \qquad (37)$$

where $\bar{V}_y$ is the average vertical velocity in the time interval $\Delta t$. Reducing the horizontal velocity within the time will be described by formula (31):

$$V_{x\,n+1} = V_{x\,n} / [1 + (C_x \rho V_{xn} * S_{tr} * \Delta t / 2m)]. \qquad (38)$$

The change in the air density with the altitude will take into account on the barometric formula $\rho = \rho_0 * \exp[-y/H_0]$, where $H_0 = 7$ km. Table 2 shows the flight parameters of the magnetic dipole depending on time. The second column shows the vertical velocity of the magnetic dipole, in the third – there is the gain altitude, in the fourth - the horizontal speed of the magnetic dipole, which it will obtain after the corresponding second of the flight.

Table 2. Dependence of the parameters on the time of flight

| t, s | $V_y$, km/s | H, km | $V_x$, km/s |
|---|---|---|---|
| 0 | 0 | 0 | 8.5 |
| 1 | 0.144 | 0.144 | 8.36 |
| 2 | 0.284 | 0.428 | 8.22 |
| 3 | 0.42 | 0.843 | 8.11 |
| 4 | 0.55 | 1.4 | 8 |
| 5 | 0.65 | 2.05 | 7.9 |
| 6 | 0.743 | 2.8 | 7.82 |
| 7 | 0.8 | 3.6 | 7.76 |
| 8 | 0.86 | 4.46 | 7.7 |
| 9 | 0.91 | 5.37 | 7.65 |
| 10 | 0.954 | 6.32 | 7.6 |
| 11 | 0.992 | 7.3 | 7.56 |
| 12 | 1.024 | 8.3 | 7.53 |
| 13 | 1.052 | 9.37 | 7.5 |
| 14 | 1.075 | 10.44 | 7.48 |

*5. Ballistics. Flight Range*

For large values of velocity $V_0 \approx 7.5$ km / s, the Earth can not be considered flat. Let us put down the equation of motion of the magnetic dipole in the cylindrical coordinate system. In this case the vertical direction will now be radial and the horizontal one - azimuthal:

$$mdV_r / dt = -mg + mV_\varphi^2/R_E, \qquad (39)$$

where $R_E = 6400$ km - the radius of the Earth, $g = 10^{-2}$ km/s$^2$ - acceleration due to gravity. Equation (39) can be reduced to:

$$dV_r / dt = -g(1 - V_\varphi^2/R_E g) = -g\,*. \qquad (40)$$

Equation (40) is $V_r = -g * t$, and, as in the case of a stone thrown at an angle to the horizon in the vacuum in the flat case, we find that the time of raising till the maximum distance and coming back to the initial point is equal to:

$$t_{max} = 2V_r / g\,*. \qquad (41)$$

For the azimuthal velocity $V_\varphi = (R_E g)^{1/2} = 8$ km / s, the time is infinite. This means that the magnetic dipole at this velocity equal to the first space velocity will rotate along a circular orbit, and will not fall back to the Earth.

For the parameters of the magnetic dipole: $V_r = 1$ km / s and $V_\varphi = 7.5$ km / s the time $t_{max}$ of lifting to the maximal height and return to the starting point is: $t_{max} = 1650$ s and, therefore, the flight range of the magnetic dipole is equal to $S_{max} = V_\varphi * t_{max} = 12300$ km. If to increase the length of the accelerator and, consequently, the finite velocity of the magnetic dipoles, their velocity after passing through the atmosphere will be more than 8 km / s and, thus, the dipoles are displayed onto near-the Earth orbit.

**Conclusion**

The magnetic moment of a current-carrying coil increases as the square of the coil that is the square of the radius growth. The coil perimeter, its mass grows with the radius increasing linearly, so that the specific magnetic moment, the magnetic moment per nucleon in the loop will grow linearly with the radius of the magnetic dipole. To limit the growth of the radius will be necessary for the sake of the sharp cone in the head part of the magnetic dipole. To have a pointed cone with the growing radius of the dipole, it will be necessary to increase its length.

You can give up using the constant magnetic field holding the magnetic dipoles against the turn by $180^0$ and alternating magnetic field keeping the magnetic dipoles near the axis of the acceleration, if the acceleration is carried out in a narrow trunk. This trunk can be made of a titanium thin walled tube with a wall thickness $\Delta h_w \approx 2$ mm. However, it is not clear how small it is necessary to choose the synchronous phase to resist the force of friction, which is not regular, between the projectile and the walls of the trunk.